# Design and numerical simulation of thermionic electron gun

M.Hoseinzade[1;1)], A.Sadighzadeh[1)]

[1] Plasma Physics and Nuclear Fusion Research School, Nuclear Science and Technology Research Institute, AEOI, PO Box 14155-1339, Tehran, Iran

**Abstract:** This paper reports the simulation of an electron gun. The effects of some parameters on the beam quality were studied and optimal choices were identified. It gives numerical beam qualities in common electrostatic triode gun, and the dependences on design parameters such as electrode geometries and bias voltages to these electrodes are shown. An electron beam of diameter 5 mm with energy of 5 keV was assumed for simulation process. Some design parameters were identified as variable parameters in the presence of space charge. These parameters are the inclination angle of emission electrode, the applied voltage to focusing electrode, the gap width between the emission electrode and the focusing electrode and the diameter of the focusing electrode. The triode extraction system is designed and optimized by using CST software (for Particle Beam Simulations). The physical design of the extraction system is given in this paper. From the simulation results, it is concluded that the inclination angle of the emission electrode is optimized at 22.5°, the applied voltage to the focusing electrode was optimized and found at Vfoc = - 600 V, also, the separation distance (gap between emission electrode and focusing electrode) = 4 mm, and the diameter of the emission electrode = 14 mm.

## 1 Introduction

Beams of charged particles and electron beams in particular are being used widely in a great and continuously increasing number of scientific instruments, electron devices and industrial facilities. An indispensable part of each electron-optical system is the electron gun in which a beam of accelerated electrons is generated. Although the electron gun is usually only a small fraction of the entire system its characteristics are crucial for the performance of the whole electron-optical column. Electron guns are routinely used for various metallurgical applications [1-5] such as melting, welding, coating, annealing, heat treatment, surface hardening, alloy formation and in atomic, molecular, and surface physics. Electron gun is also one of essential parts of electron accelerators. They are used in various types of devices such as vacuum tubes and particle accelerators. The role of electron gun is producing and shaping current of electrons in a proper form for injection into accelerating fields. Electron guns can work in a continuous or pulsed mode. There are different methods for production of burst of electrons in electron guns. Major methods are thermionic, photoelectric and electric field emissions [6]. Two important characteristics of cathode of electron guns are their emission continuity and uniformity [7, 8]. The DC thermionic electron guns basically configured by cathode, anode, emission and focusing electrodes. They have constructed in different configurations that major of them are Triode and Diode [9]. Computer simulation plays an increasingly important role in the analysis and optimization of such electron guns. It provides both detailed insight into source physics as well as the basis for improved design of high performance electron guns. Due to the importance of the triode extraction system for a high current density electron gun, the operation principle of the electron gun is carefully examined, and the triode extraction system is designed and optimized by using CST software (for Particle Beam Simulations) [10]. A detailed simulation process and the key parameters of the system are presented in this paper.

## 2 Theoretical Considerations

For simulation of electron beam guns laws and equations describing generation and emission of electrons from cathode, electric and magnetic fields and trajectories of electron beams are given in the following equations. The basic emission laws are used to describe the generation and emission of electrons. For the case when electric field extracts all the thermally generated electrons then the emission current density is given by Richardson Dushman's law,

$$J_{eT} = AT^2 \text{Exp}(\frac{-e\varphi}{KT}) \quad (1)$$

Here A is a Richardson's constant with theoretical value 120 A/cm$^2$k, while $\varphi$ is cathode work function and T is temperature of cathode.

For space charge limited emission, the emission current density of electrons is given by Child's Langmuir law as [11],

$$J = C(\frac{V^{3/2}}{d^2}) \quad (2)$$

Where C is Child's constant, V is the potential difference between the cathode and the anode lying at a distance d.

The electric field in the gun is first calculated using Poisson's equations derived from Maxwell's equations in the absence of the magnetic field. The electron trajectories are then calculated using the Lorentz force equations. The governing equations of the fields are the time-independent Maxwell's equations given by:

$$\nabla \cdot E = \frac{\rho}{\varepsilon_0} \quad (3)$$

where E is the electric field, ρ is the charge density and $\varepsilon_0$ is the permittivity of vacuum. The electron trajectories are determined by Lorentz force given by:

$$F = \frac{dP}{dt} = q(E + V \times B) \quad (4)$$

which simplify to (there is no magnetic field)

$$F = \frac{dP}{dt} = qE \quad (5)$$

Commonly, the child-Langmuir model is used for DC electron guns. [12]:

$$j_e = \frac{4\pi\varepsilon_0}{9}\sqrt{\frac{2e}{m_e}}\frac{V_0^{3/2}}{d^2} \quad (6)$$

with d in meters and $V_0$ is the applied voltage in volts. Substituting values for physical constants gives the practical expression:

$$j_e = 2.33 \times 10^{-6}\frac{V_0^{3/2}}{d^2} \quad (7)$$

The units are A/m$^2$ for d in meter.

In case of radius of < d/2, then the extraction area A = πd$^2$/4 and using eq.6, the maximum total current from an electron gun is:

$$j_e = 2.33 \times 10^{-6}\frac{\pi}{2}V_0^{3/2} \quad (8)$$

The perveance of an electron gun is defined as:

$$P = \frac{I}{V_0^{3/2}} \quad (9)$$

## 3 Thermionic electron emitter characteristics

When the filament is heated electrons can escape from the material into space. By applying a potential between the filament and the upper circular disk, the electrons can be collected. Electrons are introduced into the system by thermionic emission from the insert surface. Thermionic emission by cathodes is described by the Richardson–Dushman equation [7]:

$$J = AT^2\, e^{-e\phi/kT} \quad (10)$$

where A is, ideally, a constant with a value of 120 A/cm$^2$K$^2$, T is the temperature in kelvins, e is the charge, k is Boltzmann's constant and $\phi$ is the work function.

Temperature can be alternatively determined with Stefan-Boltzmann's law, which says that the power radiated from the surface of a hot body is:

$$P = e\sigma A\,(T^4 - T_0^4) \quad (11)$$

where P is the radiated power in watts, e is the emissivity of the material, s is the Stefan- Boltzmann constant (5.6×10$^{-8}$ W/m$^2$K$^4$), A is the surface area of the filament, T is the temperature of the filament (K) and $T_o$ is the temperature of the environment surrounding the filament.
Tungsten cathode was used for design in this study.

At high temperatures, for tungsten, the emissivity is approximately 0.35. If the filament temperature is much greater than the surrounding environment, we can neglect the $T_o$ term.
Solving for the temperature we get the expression below (good only at high temperatures)

$$T = (P/e\sigma A)^{1/4} \quad (12)$$

If we assume that the total radiated power is equal to the electrical power into the filament (P = IV), then we can calculate the temperature of the filament.
Here we use the power supply with the voltage and current of 30 V and 30 A respectively, so the power should be about 900 watt. The surface area of the filament shown in Fig.1 is roughly 2×10$^{-3}$ m$^2$. So with this power supply the temperature of the filament would be about 2200° K.

By substituting this current into eq.10, the current density emitted from this filament would be 37 mA/cm$^2$.

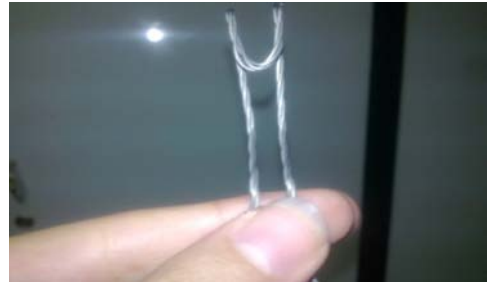

Fig. 1. Tungsten filament of electron gun

## 4 Model and description of electron gun

The electron gun is, mechanically and functionally, composed of two main parts: the electron generator and the electron beam accelerator. The electron generator is a 400 mm copper cylinder and 200 mm in diameter which is supposed to generate 25 mA current with energy of 16 kV operating in continuous mode.

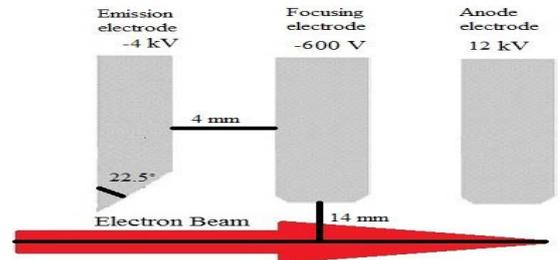

Fig. 2. Electron gun geometry elements assumed for the CST calculations

In this study, the CST particle tracking was used for simulation and precise dimensioning of the electron gun. The rough schematic of electron gun is shown in Fig. 2 and its 3D view in CST is illustrated in Fig. 3. Fig. 4 shows the equipotential lines of extraction system.

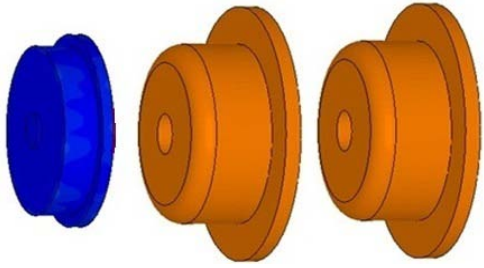

Fig. 3. Schematic of thermionic electron gun

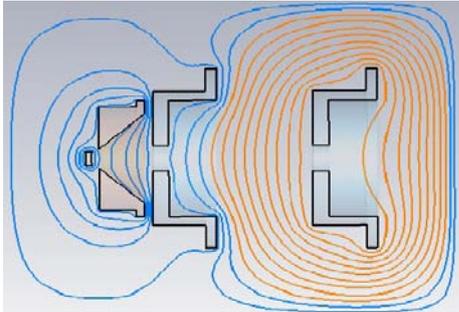

Fig. 4. (Color online) The equipotential lines of electron gun extraction system.

# 4 Simulation of thermionic electron gun by particle tracking software

Beam simulation codes have been valuable tools in understanding and designing electron gun extraction and beam transport systems. There are quite a number of different codes developed and used within this community.

The CST code [10], a program developed for simulating ion beam optics in a certain accelerator configuration, has been widely used in designing the CYCLONE30 ion source accelerator. CST PARTICLE STUDIO (CST PS) is a specialist tool for the fast and accurate analysis of charged particle dynamics in 3D electromagnetic fields. Powerful and versatile, it is suitable for tasks ranging from designing magnetrons and tuning electron tubes to modeling particle sources and accelerator components.

The particle tracking solver can model the behavior of particles through static fields, and with the gun iteration, space charge limited emission. The particle-in-cell (PIC) solver, which works in the time domain, can perform a fully consistent simulation of particles and electromagnetic fields. For relativistic applications, the wakefield solver can calculate how the fields generated by particles traveling at (or close to) the speed of light interact with the structure around them.

CST PS is integrated with the multi-purpose 3D EM modules of CST STUDIO SUITE, such as the CST EM STUDIO electro- and magnetostatic solvers and the CST MICROWAVE STUDIO eigenmode solver. It is fully embedded in the CST STUDIO SUITE design environment, thus benefitting from its intuitive modeling capabilities and powerful import interfaces. CST PS is based on the knowledge, research and development that went into the algorithms used in the MAFIA-4 simulation package. The powerful PIC solver can also make use of GPU computing, offering significant performance enhancements on compatible hardware.

Simulation of extraction systems using this simulation code is presented. The ideas applied to the accelerator design are as follows: shaping the emission electrode angle, the applied voltage to the focusing electrode, gap between emission electrode and focusing electrode, and the diameter of the focusing electrode.

## 4.1 Shape of the emission electrode

There have been long discussions over many years about what angle should be applied to the emission electrode at the corner where the electrode meets the electron beam boundary. For solid emitters such as electron guns and negative sputter ion guns the answer is the famous Pierce angle of 67.5°, which provides the coexistence of a Poisson solution inside the beam and a Laplace solution outside of it. With this condition, there are no aberrations at the beam boundary, and there is a laminar flow of charged particles. It has been accepted that shaping the emission electrode, in general, is helpful to extract a beam at a low divergence angle.

In these simulations using CST Particle Studio, the emission electrode was inclined with respect to the outer edge of the cathode at the Pierce angle of 22.5° while keeping all of the other dimensions fixed. In the configuration shown in Fig. 5, various angles including 22.5°, 30°, and 45° have been used.

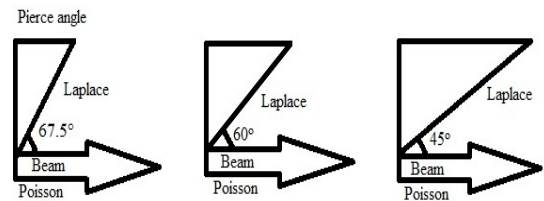

Fig. 5. Different inclination angles of emission electrode

Table 1 summarizes the simulation results. The wider the electrode angle, the more electron beam can be extracted. A wide electrode angle allows a higher field penetration into the chamber of electron gun than a small angle; therefore, more current is extracted. In addition, high field penetration into the electron gun causes distortions of the beam boundary, which leads to aberrations in the extraction system and divergence increase in the ion beam. The smallest beam diameter and emittance is achieved at small angle of 22.5◦.

Table1: extraction system simulation results with varying plasma electrode inclinations.

| Angle of inclination (degree) | Beam diameter (mm) | Emittance (cm mrad) |
| --- | --- | --- |
| 30° | 3.2° | 4. 51 |
| 22.5° | 2 ° | 4.23 |
| 45° | 4.3° | 5.35 |

## 4.2 Effect of focusing electrode voltage on beam parameters

Fig. 6 a and b show the influence of the voltage applied to the focusing electrode on both the beam emittance and beam diameter of the electron gun. It was concluded that from this figure, minimum beam emittance and beam diameter is obtained, at focusing voltage $V_{Focusing}$ of -600, -700 V, respectively.

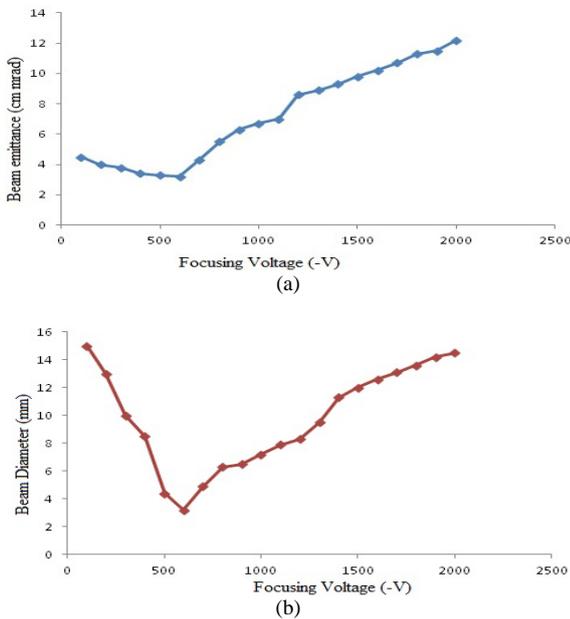

Fig. 6 a, b. Influence of the voltage applied to the focusing electrode on both beam emittance and beam diameter of the electron gun.

## 4.3 Effect of gap width of the emission and focusing electrode on beam parameters

The variation of the distance between the emission electrode and the focusing electrode (gap width) was investigated with space charge at emission voltage of -4 kV, focusing voltage of -600 V and voltage applied to the anode voltage of 12 kV (Fig.2). Fig.7 a and b shows the relation between the distance (gap width) between the emission and focusing electrode from one side and the anode electrode from other side on both the beam emittance and beam diameter of the electron gun. It was found that, minimum beam emittance for a gap width of 4 mm, whereas minimum beam diameter for a gap width of 6 mm was obtained.

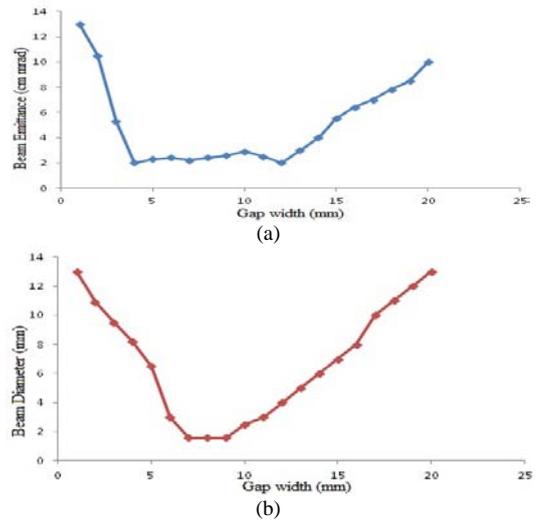

Fig. 7 a, b. Beam emittance and beam diameter as a function of the gap width of the emission and focusing electrode.

## 4.4 Effect of emission electrode diameter on beam parameters

Fig.8 a and b show the influence of gun size (inner tube diameter of the focusing electrode) on the output beam emittance and diameter. This can be attributed to the variation of the electric field inside the accelerating tube. The appropriate diameter was found to be 12 mm and 14 mm for both beam emittance and diameter, respectively of the electron gun.

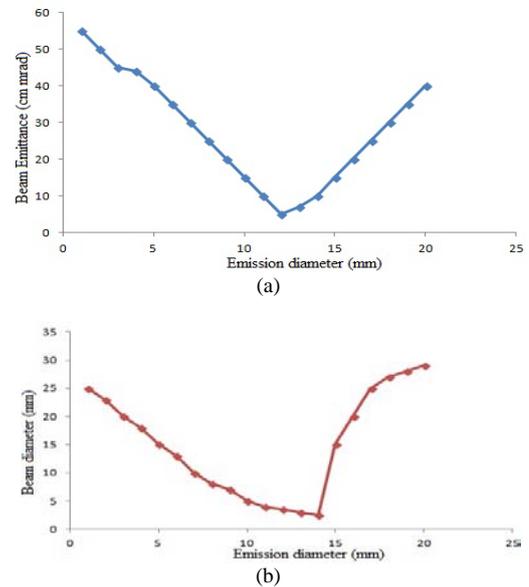

Fig. 8 a, b. Beam emittance and beam diameter as a function of the emission electrode diameter for the electron gun system.

## 5 Discussion

We could make a desirable accelerator design that satisfies our goals: low emittance and minimum beam diameter. From the above simulations, an idea has been given in regard to reforming the accelerator configuration. It was concluded that the inclination angle of emission electrode for having a good beam optic should be 22.5. Moreover, minimum beam emittance and beam diameter was found at focusing voltage of -600, -700 V, respectively. Also minimum beam emittance and beam diameter for a gap width of 4 mm and 6 mm of electron gun was obtained. The influence of gun size (inner tube diameter of the focusing electrode) on the output beam emittance and diameter was investigated. This can be attributed to the variation of the electric field inside the accelerating tube. So the appropriate diameter of emission electrode was found to be 12 mm and 14 mm for both minimum beam emittance and diameter of the electron gun respectively. Fig. 9 is the calculated profile of the extracted ion beam in the electron gun accelerator structure, which shows its acceptable beam optics properties.

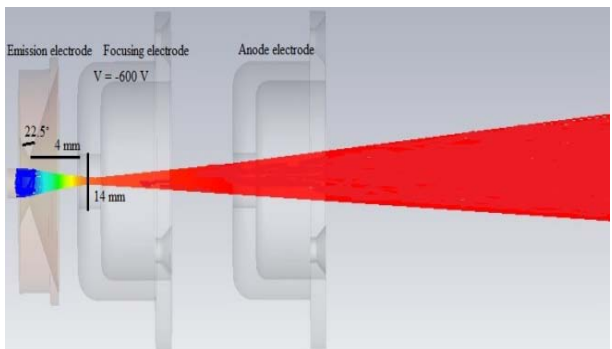

Fig. 9. Simulation run for the electron gun accelerator structure.

## 6 Conclusion

This paper presents parametric optimization and designing of an electron gun system. The influence of the voltage applied to the focusing electrode, inclination angle of emission electrode, the gap between emission and focusing electrode and the diameter of focusing electrode on both the beam emittance and beam diameter of the electron gun has been studied.

In this paper, the ion beam trajectories for different parameters of the accelerator structure were simulated and optimized. CST simulation provides some conclusions. The inclination angle of the emission electrode is optimized at 22.5°, the applied voltage to the focusing electrode was optimized and found to be roughly $V_{foc}$ = -600 V, also, the separation distance (gap between emission electrode and focusing electrode) = 4 mm, and the diameter of the emission electrode should be about 14 mm.

Also the tungsten filament was chosen as electron emitter and its temperature and current density was calculated.

The simulation results will help to direct the experimental setup that is under development at AEOI and the validity of the extraction system will be verified.